\documentclass[letterpaper,11pt]{article}

\usepackage{amsmath}
\usepackage{amsfonts}
\usepackage{amssymb}
\usepackage{graphicx}
\usepackage{color}

\usepackage{graphicx}
\usepackage[body={6.5in, 9in}, right=1in, top=1in]{geometry}
\usepackage{amssymb}
\usepackage{amsmath} 
\usepackage[numbers,sort&compress]{natbib}

\usepackage{amstext}
\usepackage{epsfig}
\usepackage{verbatim} 
\usepackage{fancybox}
\usepackage{ulem}
\usepackage{enumitem}
\usepackage{subfigure}
\usepackage{bbm}
\usepackage{parskip}
\usepackage{dsfont}
\usepackage[numbers,sort&compress]{natbib}
\usepackage{cancel}
\usepackage{pifont}
\newcommand{\cmark}{\ding{51}}%
\newcommand{\xmark}{\ding{55}}%

\def\be{\begin{equation}}
\def\ee{\end{equation}}

\newcommand\eea{\end{eqnarray}}
\newcommand\bea{\begin{eqnarray}}

\def\({\left(}
\def\){\right)}

\newcommand{\beas}{\begin{eqnarray*}}
\newcommand{\eeas}{\end{eqnarray*}}

\def\({\left(}
\def\){\right)}

\newcommand{\rd}{{\rm d}}
\newcommand{\vp}{\varphi}

\def\gsim{ \lower .75ex \hbox{$\sim$} \llap{\raise .27ex \hbox{$>$}} }
\def\lsim{ \lower .75ex \hbox{$\sim$} \llap{\raise .27ex \hbox{$<$}} }

\newcommand{\Comment}[1]{{}}
\definecolor{darkblue}{rgb}{0.15,0.35,0.55}
\definecolor{reddish}{rgb}{0.65, 0.2, 0.2}
\definecolor{darkred}{rgb}{0.7,0.3,0.3}
\definecolor{darkgreen}{rgb}{0.2,0.7,0.3}
\definecolor{darkblue2}{rgb}{0.3,0.4,0.9}
\definecolor{greyish}{rgb}{.8,.8,.8}
\usepackage[linktocpage=true]{hyperref}
\hypersetup{
colorlinks=true,
citecolor=darkblue,
linkcolor=reddish,
urlcolor=darkblue,
pdfauthor={},
pdftitle={},
pdfsubject={}
}

\usepackage{xcolor,colortbl}
\newcommand{\shade}{\cellcolor{greyish}}

\begin{document}
\def\thefootnote{\fnsymbol{footnote}}

\begin{center}
\LARGE{\textbf{From Satisfying to Violating the Null Energy Condition}} \\[0.5cm]
 
\large{Benjamin Elder$^{\rm a}$, Austin Joyce$^{\rm a,b}$ and Justin Khoury$^{\rm a}$}
\\[0.5cm]

\small{
\textit{$^{\rm a}$ Center for Particle Cosmology, Department of Physics and Astronomy, \\ University of Pennsylvania, Philadelphia, PA 19104}}

\vspace{.2cm}

\small{
\textit{$^{\rm b}$ Enrico Fermi Institute and Kavli Institute for Cosmological Physics,\\
University of Chicago, Chicago, IL 60637}}

\vspace{.2cm}

\end{center}

\vspace{.6cm}

\hrule \vspace{0.2cm}
\centerline{\small{\bf Abstract}}
\vspace{-0.2cm}
{\small We construct a theory which admits a time-dependent solution smoothly interpolating between a null energy condition (NEC)-satisfying phase at early times and a NEC-violating phase at late times. We first review earlier attempts to violate the NEC and an argument of Rubakov, presented in 1305.2614, which forbids the existence of such interpolating solutions in a single-field dilation-invariant theory. We then construct a theory which, in addition to possessing a Poincar\'e-invariant vacuum, {\it does} admit such a solution. For a wide range of parameters, perturbations around this solution are at all times stable, comfortably subluminal and weakly-coupled. The theory requires us to explicitly break dilation-invariance, so it is unlikely that the theory is fully stable under quantum corrections, but we argue that the existence of a healthy interpolating solution is quantum-mechanically robust. 
} 
\vspace{0.3cm}
\noindent
\hrule
\def\thefootnote{\arabic{footnote}}
\setcounter{footnote}{0}

\section{Introduction}

Energy conditions are usually imposed for convenience, based upon our expectations for how matter should behave. In particular, they are covariantizations of the notion that energy density should be a positive quantity.
Of all the energy conditions, the {\it Null Energy Condition} (NEC), which states that
\be
T_{\mu\nu}n^\mu n^\nu \geq 0~,
\label{nec}
\ee
for any null vector $n^\mu$,  appears to be the most fundamental. Unlike other energy conditions, it cannot be violated by the addition of a suitably large
vacuum energy contribution, so in this sense it is an unambiguous constraint on the matter. Moreover, in Einstein gravity, the NEC is necessary to establish the second law of black hole
thermodynamics~\cite{Bardeen:1973gs}. Thirdly, in cosmology the NEC precludes a non-singular bounce. Assuming spatial flatness, the Hubble parameter satisfies
\be
M_{\rm Pl}^2\dot H = -\frac{1}{2}(\rho+P)~.
\label{Hdot}
\ee
For a perfect fluid, the NEC implies $\rho+P \geq 0$, and thus $\dot{H}\leq 0$. Contraction ($H< 0$) cannot evolve to expansion ($H> 0$).
Violating~\eqref{nec} is therefore central to any alternative to inflation relying on a contracting phase before the big bang~\cite{Gasperini:1992em, Gasperini:2002bn, Gasperini:2007vw,Khoury:2001wf, Donagi:2001fs, Khoury:2001bz, Khoury:2001zk, Lyth:2001pf, Brandenberger:2001bs, Steinhardt:2001st, Notari:2002yc, Finelli:2002we, Tsujikawa:2002qc, Gratton:2003pe, Tolley:2003nx, Craps:2003ai, Khoury:2003vb, Khoury:2003rt, Khoury:2004xi, Creminelli:2004jg, Lehners:2007ac, Buchbinder:2007ad, Buchbinder:2007tw, Buchbinder:2007at, Creminelli:2007aq, Koyama:2007mg, Koyama:2007ag, Koyama:2007if, Lehners:2007wc, Lehners:2008my, Lehners:2009qu, Cai:2012va,Qiu:2011cy,Khoury:2009my, Khoury:2011ii, Joyce:2011ta, Fonseca:2011qi,Lin:2010pf,Craps:2007ch,Rubakov:2009np,Osipov:2010ee,Libanov:2010nk,Libanov:2010ci,Libanov:2011hh,Libanov:2011bk,Osipov:2013ssa,Hinterbichler:2011qk,Hinterbichler:2012mv,Koehn:2013upa, DeSantiago:2012nk}, or an expanding phase from an asymptotically static past~\cite{Nayeri:2005ck,Creminelli:2006xe,Creminelli:2010ba,LevasseurPerreault:2011mw,Liu:2011ns,Wang:2012bq,Liu:2012ww,Creminelli:2012my,Hinterbichler:2012fr,Hinterbichler:2012yn, Deffayet:2010qz}.

There also appears to be a deep, yet imprecise, connection between the NEC and well-behaved relativistic quantum field theories.
Violating the NEC generally comes hand in hand with various pathologies~\cite{Dubovsky:2005xd}, including ghosts, gradient instabilities,
superluminality, absence of a Lorentz-invariant vacuum, {\it etc}.\footnote{For example, the Hamiltonian of theories which violate the NEC was argued to be unbounded from below in~\cite{Sawicki:2012pz}.} Progress has been made in avoiding some of these shortcomings~\cite{Creminelli:2006xe,Creminelli:2010ba,Creminelli:2012my,Hinterbichler:2012fr,Rubakov:2013kaa,Hinterbichler:2012yn}, as reviewed below (see Table~\ref{NECsummary}), though a fully satisfactory example remains elusive. It is important to push this
program further, to sharpen the connection between the NEC and the standard assumptions of quantum field theory.

The {\it DBI Genesis} scenario~\cite{Hinterbichler:2012yn}, based on the DBI conformal galileons~\cite{deRham:2010eu}, is the closest any theory has come to achieving NEC violation while satisfying
the standard properties of a local quantum field theory. Specifically, as shown in~\cite{Hinterbichler:2012yn}, the coefficients of the five DBI galileon terms can be chosen such that:

\begin{enumerate}

\item The theory admits a stable, Poincar\'e-invariant vacuum. Further, the Lorentz-invariant S-matrix about this vacuum obeys the simplest dispersion relations for $2\rightarrow 2$ scattering coming from analyticity constraints.

\item The theory admits a time-dependent, homogeneous and isotropic solution which violates the NEC in a stable manner. In fact,
this NEC-violating background is an {\it exact} solution of the effective theory, including all possible higher-dimensional operators consistent with the assumed symmetries. 

\item Perturbations around the NEC-violating background, and around small deformations thereof, propagate subluminally.

\item This solution is stable against radiative corrections and the effective theory for perturbations about this solution is well-defined.

\end{enumerate}

This represents a significant improvement over ghost condensation~\cite{ArkaniHamed:2003uy} (which fails to satisfy~1) and the ordinary conformal galileons~\cite{Nicolis:2009qm, Creminelli:2010ba} (which fail to satisfy 1 and 3).\footnote{Note that the conformal galileon Lagrangian can be deformed in a straightforward way to remove superluminal propagation about the NEC-violating background~\cite{Creminelli:2012my}.} Additionally, consistency with black hole thermodynamics is desirable~\cite{Dubovsky:2006vk}. This remains an open issue which deserves further study. It is worth pointing out that the non-minimal couplings to gravity inherent in the theory will modify the usual link between NEC violation and the black hole area law. 

Unfortunately, the DBI Genesis theory itself suffers from two drawbacks. Similar to the conformal galileons, one can find weak deformations of the Poincar\'e-invariant solution around which
perturbations propagate superluminally. As pointed out recently~\cite{Creminelli:2013fxa,deRham:2013hsa}, however, galileon theories that admit superluminality can sometimes be mapped through field redefinitions to
healthy galileon theories, indicating that the apparent superluminalty is unphysical.\footnote{We thank Claudia de Rham and Andrew Tolley for a discussion on this point.} So in this sense superluminality does not offer a clear-cut criterion. But it would certainly be preferable to have an example where superluminality is manifestly absent.

A less ambiguous drawback was pointed out by Rubakov~\cite{Rubakov:2013kaa}: although the theory admits both Poincar\'e-invariant and NEC-violating solutions, any solution that attempts
to interpolate between the two vacua inevitably hits a strong coupling point. In other words, the kinetic term of fluctuations around any interpolating solution goes to zero
somewhere. In particular, it is impossible to create a NEC-violating region in the laboratory.\footnote{Although~\cite{Rubakov:2013kaa} focused on solutions which interpolate in a radial direction (a `bubble' of NEC violation), the argument applies equally well to interpolation in the temporal direction.} The argument, which we review in Sec.~\ref{rubarg}, is quite general --- it
only assumes that the theory describes a single field which is dilation-invariant, and that both Poincar\'e and NEC-violating solutions preserve this symmetry. Rubakov
showed that the conclusions can be evaded by introducing additional scalar fields.  

In this paper, we stick to a single-field theory but relax the assumption of dilation invariance in order to construct a theory which admits a solution that obeys the null energy condition at early times but at late times crosses into a phase of NEC-violation.
The theory of interest is a deformation of the
Galilean Genesis Lagrangian, 
\be
{\cal L}= {\cal Z}(\pi) e^{2 \pi} (\partial \pi)^2 + \frac{f_0^3}{\Lambda^3}(\partial \pi)^2 \Box \pi + \frac{1}{{\cal J}(\pi)}\frac{f_0^3}{2 \Lambda^3}(\partial \pi)^4~,
\label{piact1}
\ee
where the functions ${\cal Z}(\pi)$ and ${\cal J}(\pi)$ are constrained to allow a smooth interpolation between a NEC-satisfying phase at early times and a NEC-violating phase
at late times. Specifically, at early times ($\pi \rightarrow \pi_\infty$) the cubic term is negligible, and the theory reduces to
\begin{align}
\nonumber
{\cal L}_{\rm early} &\simeq -\frac{f_{\infty}^2e^{2\pi_\infty}}{\left(e^{\pi-\pi_{\infty}}-1\right)^{4}}(\partial \pi)^2 + \frac{1}{{\cal J}_0}\frac{f_0^3}{2 \Lambda^3}\frac{(\partial \pi)^4}{ \left(1 - e^{- (\pi-\pi_{\infty})}\right)^8} \\
&= - f_\infty^2e^{2\pi_\infty} (\partial\phi)^2 +\frac{f_0^3}{8{\cal J}_0\Lambda^3}(\partial\phi)^4  \,,
\label{Learly}
\end{align}
where the second line follows after a field redefinition to the (almost) canonically-normalized variable $\phi$. The quartic term has the correct sign demanded by locality~\cite{Adams:2006sv}, hence the S-matrix of this theory obeys
the standard dispersion relations coming from analyticity.

At late times ($\pi\rightarrow \infty$), on the other hand, ${\cal Z}$ and ${\cal J}$ both tend to constants, ${\cal Z}(\pi) \rightarrow f_0^2\gg \Lambda^2$, ${\cal J}(\pi) \rightarrow {\cal J}_0\sim {\cal O}(1)$, such that~\eqref{piact1} reduces to the Galilean Genesis action
\be
{\cal L}_{\rm late} \simeq f_0^2 e^{2 \pi} (\partial \pi)^2 + \frac{f_0^3}{\Lambda^3}(\partial \pi)^2 \Box \pi + \frac{1}{{\cal J}_0}\frac{f_0^3}{2 \Lambda^3}(\partial \pi)^4~.
\label{Llate}
\ee
This gives rise to the usual, genesis NEC-violating solution. For suitable values of ${\cal J}_0$, perturbations around this solution are comfortably subluminal, as in~\cite{Creminelli:2012my}.\footnote{It was recently argued that including a matter component can reintroduce superluminality in some part of the cosmological phase space~\cite{Easson:2013bda}.} Note that ${\cal Z}(\pi)$ has the correct sign at early times, and the wrong ({\it i.e.}, ghost-like) sign at late times. Nevertheless, the kinetic term of fluctuations around the
time-dependent interpolating solution is healthy during the entire evolution.

Of course, the presence of arbitrary functions in the Lagrangian makes it unlikely that the theory is radiatively stable. However, we will argue that quantum effects in the theory are under control both at early and late times. We imagine that, given this, the existence of a healthy interpolating solution is not extremely sensitive to quantum corrections, even though the functions ${\cal Z}$ and ${\cal J}$, and hence the explicit form of the solution itself, might be. In this sense, our explicit construction is designed to be a proof-of-principle. 

The paper is organized as follows. In Sec.~\ref{attempts} we briefly review earlier attempts to violate the NEC. In Sec.~\ref{rubarg} we review Rubakov's argument which forbids, in dilation invariant theories, smooth solutions interpolating between NEC-satisfying and NEC-violating vacua. After describing the construction of the theory~\eqref{piact1} in Sec.~\ref{piactconstruct}, we show in Sec.~\ref{gravneglect} that the corresponding background solution violates the NEC at late times. In Sec.~\ref{NECperts} we study perturbations around the NEC-violating solutions and derive various constraints on the parameters coming from theoretical consistency. We summarize our results and discuss future research directions in Sec.~\ref{conclu}.

\section{Attempts to violate the NEC}
\label{attempts}

In this Section, we give a brief overview of the different theories that can violate the NEC, highlighting their successes and failures. For a more comprehensive review of the null energy condition and attempts to violate it, see~\cite{Rubakov:2014jja}. In Table~\ref{NECsummary} we provide a scorecard for the different theories. A natural place to search for matter which can violate the NEC is in the context of scalar field theories, since scalars can develop nontrivial background profiles that preserve homogeneity and isotropy.

\begin{itemize}

\item {\bf 2-derivative theories:} Consider a non-linear sigma model with dynamical variables $\phi^I: {\mathbb R}^{3,1} \to {\cal M}$, where ${\cal M}$ is an arbitrary, $N$-dimensional real target space. At 2-derivative order, the action is given by\footnote{Throughout, we use the mostly plus $(-, +,+,+)$ metric convention.}
\be
S =\int\rd^4 x\left(- \frac{1}{2} G_{I J}(\phi) \partial_\mu \phi^I \partial^\mu \phi^J-V(\phi^I\phi_I)\right)~,
\ee
where $G_{IJ}$ is the target-space metric. The stress-energy tensor for this field is readily computed, and the quantity relevant for the NEC is
\be
T_{\mu\nu}n^\mu n^\nu = G_{IJ}(\phi) n^\mu n^\nu \partial_\mu\phi^I  \partial_\nu\phi^J\,.
\ee
In the language of perfect fluids, focusing on time-dependent profiles, $\phi^I(t)$, this becomes
\be
\rho+P = G_{IJ}(\phi)\dot{\phi}^I\dot{\phi}^J~.
\ee
Now, the target-space metric $G_{IJ}$ can be diagonalized, since it is symmetric and invertible. Therefore, in order to violate the NEC ($\rho+P \leq 0$), $G_{IJ}$ must have at least one negative eigenvalue,
that is, one of the $\phi^I$'s must be a ghost. At the 2-derivative level, violating the NEC comes hand in hand with ghosts.

\item {\bf $P(X)$ theories:} The obvious generalization is to consider higher-derivative theories. In order to avoid ghost instabilities, the equation of motion should remain $2^{\rm nd}$-order. A general class of such models is
\be
S = M^4\int\rd^4x P(X)~,
\label{pXlag}
\ee
where $X\equiv -\frac{1}{2M^4}(\partial\phi)^2$, and $M$ is an arbitrary mass scale. The justification for considering theories of this type is effective field theory reasoning --- we anticipate that at low enough energies, terms with more derivatives per field will be sub-leading. However, even in these theories, NEC violation generically introduces pathologies, albeit of a more subtle nature. 

To see this, note that the combination $\rho+P$, assuming $\phi = \phi(t)$, is given by
\be
\rho+P = 2XP_{,X}~.
\ee
In order to violate the NEC, we therefore need $P_{,X} < 0$. Meanwhile, expanding~\eqref{pXlag} about the background $\phi = \bar\phi(t)+\vp$, the action for quadratic fluctuations is~\cite{ArkaniHamed:2003uy} 
\be
S_\vp = \frac{1}{2} \int\rd^4x \left(\left(P,_{X}+2XP,_{XX}\right)\dot\vp^2-P,_{X}(\vec\nabla\vp)^2\right)~.
\label{pXvplag}
\ee
A violation of the NEC ($P,_{X} < 0$) results in either gradient instabilities (wrong-sign spatial gradient term) or ghost instabilities (if we choose $P,_{X}+2XP,_{XX} < 0$). More generally, it was shown in~\cite{Dubovsky:2005xd,Buniy:2006xf} that violating the NEC in theories of the form ${\cal L}(\phi^I, \partial\phi^I)$ ({\it i.e.}, involving at most one derivative per field), implies either the presence of ghost or gradient instabilities {\it or} superluminal propagation.

\item {\bf Ghost condensation:} This general theorem about instabilities in such a wide class of theories would seem to preclude any sensible violations of the NEC. There is, however, a rather compelling loophole to the general logic. The theorem of~\cite{Dubovsky:2005xd} relies heavily on the standard organization of effective field theory, {\it i.e.}, the sub-dominance of terms of the form $\partial^2\phi$. There exist two well-studied situations where such terms can become important and, indeed, both lead to violations of the NEC free of the obvious pathologies. 

The first is {\it ghost condensation}~\cite{ArkaniHamed:2003uy}. This relies on an action of the $P(X)$ form~\eqref{pXlag}, but chosen so that there exists a solution with $P,_{X} = 0$.
Notice from~\eqref{pXvplag} that this precisely corresponds to the vanishing of the spatial gradient term in the quadratic Lagrangian about this background. This allows a higher-derivative term of the form $(\nabla^2\vp)^2$ to become important in the quadratic Lagrangian {\it without} the effective field theory expansion breaking down. Since $\rho + P = 0$ on the background, this acts as a vacuum energy contribution. Deforming the background as $\phi = \bar{\phi}(t) +\pi(t)$, one finds
\be
\rho + P \sim \dot{\pi}\,.
\ee
This is {\it linear} in $\pi$, and hence can have either sign. Violating the NEC once again will push the kinetic term of fluctuations slightly negative, but the dispersion relation is stabilized at high $k$ by the $(\nabla^2\vp)^2$ term~\cite{Creminelli:2006xe}. The no-go theorem of~\cite{Dubovsky:2005xd} is thus evaded by relying on higher-derivative spatial gradient terms. 

The main drawback of the ghost condensate is the absence of a Lorentz-invariant vacuum. Indeed, from~\eqref{pXvplag} the absence of ghosts about the condensate $P_{,X} = 0$ solution requires $P_{,XX} > 0$,
{\it i.e.}, the condensate is at a minimum of $P(X)$. As a result, the theory cannot be connected to a Lorentz-invariant vacuum ($P_{,X}\vert_{X = 0} > 0$) without encountering pathologies in between. The theory is only well-defined in the neighborhood of the ghost condensate point. 

A NEC-violating ghost condensate phase has been used in alternative cosmological models, including a universe starting from an asymptotically static past~\cite{Creminelli:2006xe}, the New Ekpyrotic Universe~\cite{Buchbinder:2007ad,Creminelli:2007aq}, and the matter-bounce scenario~\cite{Lin:2010pf}.

\item {\bf Galileons:} A second class of theories which can violate the NEC without instabilities is given by the conformal {\it galileons}~\cite{Nicolis:2008in,Nicolis:2009qm}.\footnote{For another construction which violates the NEC based on Kinetic Gravity Braiding~\cite{Deffayet:2010qz}, a cousin of the galileons, see~\cite{Easson:2011zy}.} These are conformally-invariant scalar field theories with derivative interactions. The simplest example is
\be
{\cal L} = f^2e^{2\pi}(\partial\pi)^2+\frac{f^3}{\Lambda^3}\square\pi(\partial\pi)^2+\frac{f^3}{2\Lambda^3}(\partial\pi)^4~.
\label{galileonoriginal}
\ee
Each term is manifestly dilation invariant. The relative $1/2$ coefficient between the $\square\pi(\partial\pi)^2$ and $(\partial\pi)^4$ ensures
full conformal invariance.\footnote{Under the dilation and conformal symmetries, the field $\pi$ transforms as:
\be
\delta_D\pi = -1-x^\mu\partial_\mu\pi~;~~~~~~~~~~\delta_{K^\mu}\pi = -2x^\mu-(2x^\mu x^\nu\partial_\nu-x^2\partial^\mu)\pi~.
\ee
}
Choosing the kinetic term to have the wrong sign, as in~\eqref{galileonoriginal}, the theory admits a time-dependent solution
\be
e^{\pi} = \frac{1}{H_0(-t)}~;~~~~~~H_0^2=\frac{2}{3}\frac{\Lambda^3}{f}~,
\label{genback}
\ee
where $-\infty < t <0$. For consistency of the effective field theory, the scale $H_0$ should lie below the strong coupling scale $\Lambda$, which requires
\be
f\gg \Lambda\,.
\ee
This background spontaneously breaks the original SO$(4,2)$ symmetry down to its SO$(4,1)$ subgroup. The stress-energy violates the NEC~\cite{Creminelli:2010ba,Nicolis:2009qm, Creminelli:2012my}: $\rho + P = - \frac{2f^2}{H_0^2t^2}$. Perturbations around this solution are stable, and propagate exactly luminally by SO$(4,1)$ invariance. However, the sound speed can be pushed to superluminal values on slight deformations of this background.\footnote{It was shown in~\cite{Easson:2013bda} that such deformations must break homogeneity/isotropy.} A cure to this pathology~\cite{Creminelli:2012my} is to reduce the symmetry by detuning the relative coefficient of the cubic and quartic terms
\be
{\cal L} = f^2e^{2\pi}(\partial\pi)^2+\frac{f^3}{\Lambda^3}\square\pi(\partial\pi)^2+\frac{f^3}{2\Lambda^3}(1+\alpha)(\partial\pi)^4~,
\label{sublum}
\ee
where $\alpha$ is a constant. For $\alpha \neq 0$, this explicitly breaks the special conformal symmetry, leaving dilation and Poincar\'e transformations as the only symmetries (which conveniently close to form a subgroup). This still allows a $1/t$ background of the form~\eqref{genback}, with $H_0 = \frac{2}{3}\frac{1}{(1+\alpha)}\frac{\Lambda^3}{f}$ depending on $\alpha$. For $-1 < \alpha < 3$, this background violates the NEC and has stable perturbations. As a result of the fewer residual symmetries, perturbations propagate with a sound speed different from unity:
\be
c_s^2 = \frac{3-\alpha}{3(1+\alpha)}\,.
\ee
This is subluminal for $\alpha > 0$. In other words, for the range
\be
0 < \alpha < 3\,,
\ee
the system violates the NEC, is stable against small perturbations, and these perturbations propagate at subluminal speeds.
Moreover, the theory is stable against quantum corrections.

The main drawback of the galileon NEC violation is --- just like the ghost condensate --- the absence of a Lorentz-invariant vacuum. Indeed, the existence and stability of a $1/t$ background requires a wrong-sign kinetic term, as in~\eqref{galileonoriginal} and~\eqref{sublum}. As shown in~\cite{Creminelli:2012my}, including the higher-order conformal galileon terms does not help: only the kinetic term contributes to $\rho + P$ for the $1/t$ solution, and it must have the wrong sign to violate NEC. An improvement over the ghost condensate, however, is that perturbations are stable on all scales, whereas perturbations of the ghost condensate in the NEC-violating phase are unstable on large scales (but are stabilized on small scales, thanks to the higher-derivative contribution to the dispersion relation).

A NEC-violating galileon phase is the hallmark of the Galilean Genesis scenario~\cite{Creminelli:2010ba,Creminelli:2012my}, in which the universe expands from an asymptotically static past. 
Because of the residual dilation symmetry, nearly massless fields acquire a scale invariant spectrum. The SO$(4,2)\rightarrow$ SO$(4,1)$ spontaneous breaking is also used in the
NEC-satisfying rolling scenario of~\cite{Craps:2007ch,Rubakov:2009np,Hinterbichler:2011qk}. More generally, this symmetry breaking pattern arises whenever a number of scalar operators ${\cal O}_I$ with weight $\Delta_I$ in a conformal field theory acquire a time-dependent profile ${\cal O}_I(t) \sim (-t)^{-\Delta_I}$. The general effective action was constructed in~\cite{Hinterbichler:2012mv} utilizing the coset construction, and the consistency relations were derived in~\cite{Creminelli:2012qr}.

\item {\bf DBI Galileons:} An alternative way to avoid superluminality while preserving the full SO$(4,2)$ symmetry is to consider the DBI conformal galileons~\cite{deRham:2010eu}. These are the ``relativistic" extension of the ordinary conformal galileons, and describe the motion of a 3-brane in an AdS$_5$ geometry. The DBI conformal galileon action is a sum of five geometric invariants, with 5 free coefficients $c_1,\ldots, c_5$:
\be
{\cal L} = c_1{\cal L}_1  + c_2 {\cal L}_2 + c_3 {\cal L}_3 + c_4 {\cal L}_4 + c_5 {\cal L}_5\,,
\ee
where the ${\cal L}_i$'s are built out of the induced metric 
\be
\bar g_{\mu\nu} = G_{AB} \partial_\mu X^A \partial_\nu X^B  = \phi^2\left( \eta_{\mu \nu} + \frac{\partial_\mu \phi\partial_\nu \phi}{\phi^4}\right) \,,
\ee
the Ricci tensor $\bar R_{\mu\nu}$ and scalar $\bar R$, and the extrinsic curvature tensor
\be
K_{\mu\nu} = \gamma \phi^2 \left(\eta_{\mu\nu} -\frac{\partial_\mu\partial_\nu\phi}{\phi^3} + 3\frac{\partial_\mu\phi\partial_\nu\phi}{\phi^4}\right)\,.
\ee
Each ${\cal L}_i$ is invariant up to a total derivative under SO$(4,2)$ transformations, inherited from the isometries of AdS$_5$. The relevant terms come from considering brane Lovelock invariants~\cite{Lovelock:1971yv} and the boundary terms associated to bulk Lovelock invariants:
\begin{align}
\nonumber
{\cal L}_1 &= -\frac{1}{4} \phi^4\,;  \\
\nonumber
{\cal L}_2 &= -\sqrt{-\bar g} = - \frac{\phi^4}{\gamma}\,; \\
\nonumber
{\cal L}_3 &= \sqrt{-\bar g} K = - 6 \phi^4 + \phi [\Phi] + \frac{\gamma^2}{\phi^{3}} \Big( - [\phi^3] + 2 \phi^7 \Big)\,;\\
\nonumber
{\cal L}_4 &= - \sqrt{-\bar g} \bar R \\
\nonumber
&= 12 \frac{\phi^4}{\gamma} + \frac{\gamma}{\phi^{2}} \Big( [\Phi^2] - \left( [\Phi] - 6 \phi^3 \right) \left( [\Phi] - 4 \phi^3 \right) \Big) + 2 \frac{\gamma^3}{\phi^{6}} \Big(  - [\phi^4] + [\phi^3] \left( [\Phi] - 5 \phi^3 \right) - 2 [\Phi] \phi^7 + 6 \phi^{10} \Big)\,;\\
\nonumber
{\cal L}_5 &= \frac{3}{2} \sqrt{-\bar g}   \left(- \frac{K^3}{3}  + K_{\mu \nu}^2 K - \frac{2}{3} K_{\mu \nu}^3 - 2 \left(\bar R_{\mu \nu}-\frac{1}{2}\bar R\bar g_{\mu\nu}\right)  K^{\mu \nu}\right) \\
\nonumber
&= 54 \phi^4 - 9 \phi [\Phi]  + \frac{\gamma^2}{\phi^{5}} \bigg( 9 [\phi^3] \phi^2 + 2 [\Phi^3] - 3 [\Phi^2] [\Phi] + 12 [\Phi^2] \phi^3 + [\Phi]^3 - 12 [\Phi]^2 \phi^3 + 42 [\Phi] \phi^6 - 78 \phi^4 \bigg)  \\
\nonumber
&~~~~~~~~+    3 \frac{\gamma^4}{\phi^{9}} \bigg( -2 [\phi^5] + 2 [\phi^4] \left( [\Phi] - 4 \phi^3 \right) + [\phi^3] \left( [\Phi^2] - [\Phi]^2 + 8 [\Phi] \phi^3 - 14 \phi^6 \right) \\
\nonumber
&  ~~~~~~~~~~~~~~~ + 2 \phi^7 \left( [\Phi]^2 - [\Phi^2] \right) - 8 [\Phi] \phi^{10} + 12 \phi^{13} \bigg) \,,
\label{Ls}
\end{align}
where $\gamma\equiv 1/\sqrt{1 + (\partial\phi)^2/\phi^4}$ is the Lorentz factor for the brane motion, ${\cal L}_1$ measures the proper 5-volume between the brane and some fixed
reference brane ${\cal L}_2$ is the world-volume action~\cite{Goon:2011qf}, {\it i.e.}, the brane tension, and the higher-order terms ${\cal L}_3$, ${\cal L}_4$ and ${\cal L}_5$ are
various functions of curvature. Moreover, $\Phi$ denotes the matrix of second derivatives $\partial_\mu\partial_\nu \phi$, $[\Phi^n] \equiv {\rm Tr}(\Phi^n)$, and $[\phi^n] \equiv \partial\phi \cdot \Phi^{n-2} \cdot \partial\phi$, with indices raised by $\eta^{\mu\nu}$. The motivation for considering Lovelock terms is that they lead to second-order equations of motion for the scalar field $\phi$~\cite{deRham:2010eu}.

For suitable choices of the coefficients $c_1,\ldots, c_5$, the theory admits a $1/t$ solution of the form~\eqref{genback}, which violates the NEC in a stable manner~\cite{Hinterbichler:2012yn}.
This was dubbed the {\it DBI Genesis} phase in~\cite{Hinterbichler:2012yn}. Analogous to DBI inflation~\cite{Alishahiha:2004eh}, the sound speed of fluctuations for relativistic brane motion $\gamma \gg 1$ is highly subluminal. This is an improvement over the galileon examples, since subluminality is achieved while keeping the full conformal symmetries. Moreover, this solution is stable against radiative corrections:
terms not of the conformal DBI form are generated radiatively but with coefficients suppressed by inverse powers of $\gamma$.

More importantly, the theory also admits a stable, Poincar\'e-invariant vacuum. As such, DBI Genesis is the first example of a theory possessing both stable NEC-violating and stable Poincar\'e-invariant vacua.
In~\cite{Hinterbichler:2012yn} it was shown that the $2\to 2$ scattering amplitude satisfies the known analyticity properties required by locality. Unfortunately, like ordinary galileons weak-field deformations 
of the Poincar\'e-invariant vacuum allow superluminal propagation of perturbations. Hence we naively do not expect the full scattering S-matrix to be analytic, though as mentioned earlier it is not clear to what
extent the apparent superluminality is truly a pathology~\cite{Creminelli:2013fxa,deRham:2013hsa}.

\end{itemize}

\begin{table}[float]
\centering
\small
\begin{tabular}{| l | c | c | c | c |}
	\hline
	& {\bf Ghost condensate} & {\bf Galilean Genesis} & {\bf DBI Genesis} & {\bf This theory} \\ \hline
	{\bf \cancel{NEC} vacuum} &  {\color{darkgreen}\cmark} & {\color{darkgreen}\cmark} &  {\color{darkgreen}\cmark} &  {\color{darkgreen}\cmark}\\		\hline
	 No ghosts &  {\color{darkgreen}\cmark} &  {\color{darkgreen}\cmark} &  {\color{darkgreen}\cmark} &  {\color{darkgreen}\cmark} \\\hline
	  Sub-luminality &  {\color{darkgreen}\cmark} &  {\color{darkgreen}\cmark} &  {\color{darkgreen}\cmark} &  {\color{darkgreen}\cmark} \\\hline
	{\bf Poincar\'e vacuum} & {\color{darkred}\xmark} & {\color{darkred}\xmark} & {\color{darkgreen}\cmark} & {\color{darkgreen}\cmark} \\ \hline
	No ghosts &  \shade  &  \shade& {\color{darkgreen}\cmark} & {\color{darkgreen}\cmark}\\		\hline
	S-Matrix analyticity ($2\to2$) &   \shade &  \shade& {\color{darkgreen}\cmark} &  {\bf {\color{darkgreen} \cmark}}\\\hline
	Sub-luminality & \shade&\shade  & {\color{darkred}\xmark} & {\color{darkgreen}\cmark}\\		\hline
	{\bf Interpolating solution}  &  \shade &  \shade & {\color{darkred}\xmark} & {\color{darkgreen}\cmark}  \\ \hline
	{\bf Radiative stability}  & {\color{darkgreen}\cmark} & {\color{darkgreen}\cmark} & {\color{darkgreen}\cmark} & {\color{darkred}\xmark} \\ \hline
	{\bf BH Thermodynamics}  & {\color{darkred}\xmark} & {\bf {\color{darkblue2} ?}} & {\bf {\color{darkblue2} ?}} & {\bf {\color{darkblue2} ?}}  \\ \hline
\end{tabular}
\caption{\small Checklist of properties of various theories which possess null energy condition-violating solutions.}
\label{NECsummary}
\end{table}

Additionally, one would like the theory to be consistent with the second law of black hole thermodynamics. There appears to be great tension between this and the NEC, for instance in the ghost condensate violation of the NEC allows for the formation of perpertuum mobile~\cite{Dubovsky:2006vk}.\footnote{For a contrary viewpoint, see~\cite{Mukohyama:2009rk}.} The story is potentially more subtle for
DBI galileons, thanks to the non-minimal terms required for their covariantization~\cite{Deffayet:2009wt, deRham:2010eu}. This is currently under investigation~\cite{yizen}.

Ideally, one would like to be able to start from the Poincar\'e-invariant vacuum and evolve smoothly into the NEC-violating phase. 
As pointed out recently~\cite{Rubakov:2013kaa}, however, this is impossible in any single-field theory with dilation invariance. This is particularly constraining because many of the attempts to violate the NEC utilize dilation-invariant theories (for example the Galilean Genesis scenarios and the DBI conformal galileons). The argument, reviewed below, shows that any solution that attempts to interpolate between the two vacua inevitably hits a strong coupling point.
One way out is to invoke multiple scalar fields. Another way out, which we will explore here, is to break the dilation symmetry explicitly. In doing so, we will be
able to construct a theory with the following properties:
\begin{itemize}
\item A Poincar\'e-invariant vacuum with stable and sub-luminal fluctuations about this vacuum.
\item A solution which interpolates between a non-NEC-violating phase and a phase of NEC violation with stability and sub-luminality for perturbations about this solution.
\end{itemize}

\section{A no-go argument for interpolating solutions}
\label{rubarg}

In this Section we review the no-go argument of Rubakov~\cite{Rubakov:2013kaa}, which forbids the existence of a well-behaved solution interpolating between {\it dilation-invariant} vacua. 
The argument is very general and applies to any single scalar field theory that enjoys (at the classical level) dilation invariance, and admits both a Poincar\'e invariant solution
and a dilation-preserving, NEC-violating background.

First note that conservation of the energy-momentum tensor is equivalent to the equation of motion via
\be
\partial_\mu T^{\mu}_{~\nu} = -\frac{\delta S}{\delta\pi} \partial_\nu\pi~,
\ee
where $\delta S/\delta\pi$ is the Euler--Lagrange derivative. Specializing to $\pi = \pi(t)$, it follows the equation of motion is equivalent to energy conservation:
\be
\dot{\rho} = - \dot{\pi} \frac{\delta S}{\delta\pi}~.
\label{phiEOM}
\ee
Now we assume that the equation of motion is second-order, that is, $\delta S/\delta\pi$ contains at most $\ddot{\pi}$ but no higher-derivatives.\footnote{Note that violating this assumption would lead to Ostrogradski-type instabilities~\cite{Ostrogradski, Woodard:2006nt}.}
It then follows that $\rho$ must be a function only of $\pi$ and $\dot\pi$, for otherwise $\dot\rho$ would contribute higher-derivative terms in~\eqref{phiEOM}.
Since the theory is dilation invariant, we can deduce the form of $\rho$:\footnote{Under a finite dilation, $x^\mu\mapsto \lambda x^\mu$, the field $\pi$ transforms as $\pi(x) \mapsto \pi(\lambda x)+\log \lambda$. One can then check that~\eqref{rhoexplicit} is the most general object depending only on $\pi$ and $\dot\pi$ invariant under this symmetry.}
\be
\rho = e^{4\pi} Z\left(e^{-2\pi} \dot\pi^2\right)~,
\label{rhoexplicit}
\ee
where $Z$ is a theory-dependent function. 

If the theory admits a Poincar\'e-invariant solution, $\pi = {\rm constant}$, it will have vanishing energy density:
\be
Z(0) = 0\,.
\ee
Additionally, if the theory admits a NEC-violating background which preserves homogeneity and isotropy, then $\pi$ can only depend on time.
If this background is also dilation invariant, then it must take the form $e^\pi \sim t^{-1}$, and hence  $e^{-2\bar{\pi}} \dot{\bar{\pi}}^2 \equiv Y = {\rm constant}$
on this solution. Moreover, the assumed symmetries imply $\rho = \beta t^{-4}$ on the time-dependent solution, while energy conservation requires $\dot{\rho} =0$, and thus $\beta = 0$. 
It follows that
\be
Z(Y) = 0\,.
\ee
In other words, the energy density vanishes on any background that preserves homogeneity, isotropy and dilation symmetry. This of course includes the Poincar\'e-invariant vacuum
and (by assumption) the NEC-violating background.

\begin{figure}
\centering
\includegraphics[width=3.5in]{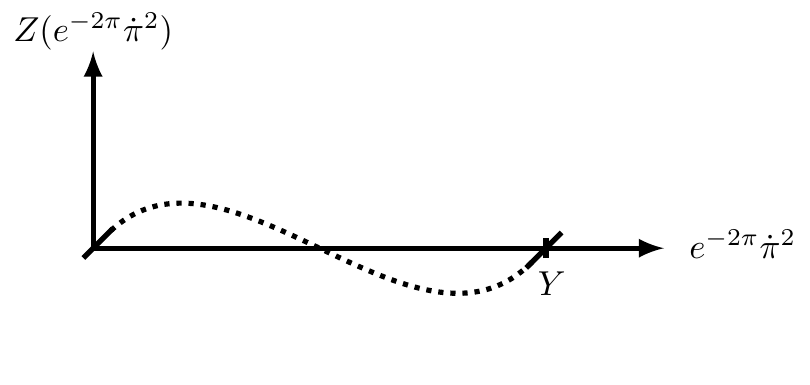}
\caption{\small In a dilation invariant theory, we must have $Z = 0$ at both $e^{-2\pi} \dot\pi^2 =0$ and $e^{-2\pi} \dot\pi^2= Y$, as well as $Z' > 0$ at both of these points. It is impossible to connect these two solutions without having a region where $Z' <0$, as is clear from the plot.}
\label{rubdiag}
\end{figure}

Next consider the stability of these solutions. We can expand~\eqref{phiEOM} about some time-dependent background, $\pi = \bar\pi(t)+\varphi$, and use the form~\eqref{rhoexplicit} for $\rho$ to derive an equation of motion for $\varphi$. For the diagnosis of ghost instabilities, we only explicitly need the $\ddot\varphi$ term:
\be
-2e^{2\bar\pi}\bar{Z}' \ddot\varphi +\cdots = 0~.
\ee
This clearly derives from the quadratic Lagrangian ${\cal L} =  e^{2\bar\pi}\bar{Z}' \dot\varphi^2+\cdots$. In order for $\varphi$ to be healthy, we must have
\be
\bar{Z}' > 0\,.
\ee
The problem is now clear: we have two backgrounds, each with $Z = 0$. In order for them to both be healthy, we must have $Z' > 0$ around each solution. 
 It is impossible to connect these two solutions without having $Z' < 0$ --- and hence developing a ghost --- somewhere in between. More physically, in trying
 to connect these solutions one must inevitably hit $Z = 0$, which corresponds to strong coupling. It is therefore impossible to connect the two backgrounds with
 a solution which is perturbative. See Fig.~\ref{rubdiag} for a graphical representation of this result.

This no-go argument is very general, but we can get some inspiration for how to avoid it by examining its assumptions. The most natural ones to consider breaking are the assumption of a single degree of freedom and that of dilation invariance. Indeed, Rubakov considers a model which introduces additional degrees of freedom to construct an interpolating solution~\cite{Rubakov:2013kaa}. Here, we will focus on theories that are not dilation invariant.

\section{Construction of the theory}
\label{piactconstruct}

To circumvent the no-go argument of Sec.~\ref{rubarg}, we stick to a single-field theory but relax the assumption of dilation invariance. We consider a deformation of the conformal galileon lagranagian~\eqref{galileonoriginal} (used in Galilean Genesis~\cite{Nicolis:2009qm,Creminelli:2010ba}) by introducing functions ${\cal Z}(\pi), {\cal J}(\pi)$ which explicitly break scale invariance:
\be
{\cal L}= {\cal Z}(\pi) e^{2 \pi} (\partial \pi)^2 + \frac{f_0^3}{\Lambda^3}(\partial \pi)^2 \Box \pi + \frac{1}{{\cal J}(\pi)}\frac{f_0^3}{2 \Lambda^3}(\partial \pi)^4~.
\label{piact}
\ee
Our goal is to find suitable functional forms for ${\cal Z}(\pi)$ and ${\cal J}(\pi)$ such that the theory admits a smooth solution which is NEC-satisfying at early times ($t\ll t_*$), and NEC-violating at late times ($t\gg t_*$). The transition time will be denoted by $t_*$.

\subsection{Late time behavior}

To achieve NEC violation with strictly subluminal propagation of perturbations at late times ($t\gg t_*$), the theory should approximate the form~\eqref{sublum}, used in subluminal genesis~\cite{Creminelli:2012my}. This requires 
\be
{\cal Z}(\pi) \to f_0^2\,;~~~{\cal J}(\pi) \to {\cal J}_0~~~~~~{\rm for}~~~t \gg t_*\,,
\label{latecond}
\ee
where $f_0 \gg \Lambda$ and ${\cal J}_0$ is an ${\cal O}(1)$ constant. Thus, the theory reduces at late times to
\be
{\cal L}_{\rm late} \simeq f_0^2 e^{2 \pi} (\partial \pi)^2 + \frac{f_0^3}{\Lambda^3}(\partial \pi)^2 \Box \pi + \frac{1}{{\cal J}_0}\frac{f_0^3}{2 \Lambda^3}(\partial \pi)^4~.
\label{Llatemain}
\ee
Comparison with~\eqref{sublum} gives the translation
\be
{\cal J}_0 = \frac{1}{1 + \alpha}\,,
\label{jalpha}
\ee
hence we anticipate that we will need ${\cal J}_0~\lsim~{\cal O}(1)$ to have subluminality~\cite{Creminelli:2012my}. At late times, the solution should therefore asymptote to the Genesis background
\be
e^{\pi} = \frac{1}{H_0(-t)}~;~~~~~~H_0^2=\frac{2{\cal J}_0}{3}\frac{\Lambda^3}{f_0}~~~~~~{\rm for}~~~t \gg t_*\,.
\label{pibackmain}
\ee
The energy scale of this solution is $H_0$. We demand that it lie below the strong coupling scale of the effective theory, $H_0\ll \Lambda$, which will be the case if
\be
f_0 \gg \Lambda\,.
\label{flambcond}
\ee
The background~\eqref{pibackmain} is a solution on flat, Minkowski space. With gravity turned on, it remains an approximate solution at early enough times in the Genesis phase.
Gravity eventually becomes important at a time $t_{\rm end}$, which will be computed in Sec.~\ref{gravneglect}.

\subsection{Early time behavior}

At early times ($t \ll t_*$), the solution should asymptote to a constant field profile:
\be
\pi \simeq \pi_{\infty}~~~~~{\rm for}~~~~~t \ll t_*\,.
\ee
In order for this constant background to be ghost-free, the sign of the kinetic term should be the usual (negative) one:
\be
{\cal Z}(\pi) < 0~~~~~{\rm for}~~~~~t \ll t_*\,.
\label{earlycond}
\ee
We will see that this gives rise to a NEC-satisfying phase, with $\rho \sim P$. In this regime, clearly gravity cannot be ignored arbitrarily far in the past,
since the universe must emerge from a big bang singularity. We will come back to this point in Sec.~\ref{gravneglect} and show that the time $t_{\rm beg}$ 
where gravity becomes important is parametrically larger in magnitude than $t_*$. In other words, there is a parametrically large window $t_{\rm beg} \ll t\ll t_*$
within which gravity is negligible and the early-time expressions above hold. In particular, $\pi$ can be approximated as constant over this regime, in the sense that
it varies slowly compared to the Hubble parameter at the transition. 

From~\eqref{latecond} and~\eqref{earlycond}, note that ${\cal Z}(\pi)$ has the correct sign at early times, and the wrong ({\it i.e.}, ghost-like) sign at late times (as required for the Genesis solution).
Nevertheless, we will see that the kinetic term of fluctuations around the time-dependent solution is healthy during the entire evolution. This does imply, however, that stable, Lorentz-invariant
vacua only exist for a finite range in field space.

\subsection{Interpolating functions}

We engineer the desired ${\cal Z}(\pi)$ and ${\cal J}(\pi)$ by demanding that they give rise to a suitable time-dependent background solution,
which interpolates between $\pi \simeq \pi_{\infty}$ at early times and $e^\pi \sim 1/t$ at late times. A simple ansatz for the background
which satisfies these asymptotic conditions is
\be
e^{\bar{\pi}(t)} = e^{\pi_{\infty}} \left(1 + \frac{t_*}{t}\right) \,,
\label{piansatz}
\ee
where $t_*$ sets the transition time.

Assuming spatial homogeneity, the equation of motion for $\pi = \bar{\pi}(t)$ following from~\eqref{piact} is
\be
\ddot{\bar{\pi}} \left(-{\cal Z}(\bar{\pi}) e^{2\bar{\pi}}  + \frac{3}{{\cal J}(\bar{\pi})} \frac{f_0^3}{\Lambda^3}\dot{\bar{\pi}}^2\right) = \frac{\dot{\bar{\pi}}^2}{2} \left(\left( 2{\cal Z}(\bar{\pi})+ {\cal Z}'(\bar{\pi}) \right) e^{2\bar{\pi}} + \frac{3 {\cal J}'(\bar{\pi})}{2 {\cal J}^2(\bar{\pi})} \frac{f_0^3}{\Lambda^3} \dot{\bar{\pi}}^2\right)\,.
\label{pieom1st}
\ee
This admits a first integral of motion enforcing energy conservation:
\begin{align}
\nonumber
\rho &= -{\cal Z}(\bar\pi) e^{2 \bar{\pi}} \dot{\bar{\pi}}^2 + \frac{3}{2 {\cal J}(\bar{\pi})} \frac{f_0^3}{\Lambda^3} \dot{\bar{\pi}}^4 \\
&=  -{\cal Z}( t ) e^{2\pi_{\infty}} \frac{t_*^2}{t^4} + \frac{3}{2 {\cal J}( t )} \frac{f_0^3}{\Lambda^3} \frac{t_*^4}{t^4 (t + t_*)^4} = {\rm constant}\,,
\label{rhocons}
\end{align}
where in the second line we have substituted in the background solution~\eqref{piansatz}.
At early times ($t\ll t_*$), the two contributions scale differently: $\sim t^{-4}$ for the first term; $\sim t^{-8}$ for the second.
The simplest option is for each term to be separately constant, from which we can deduce the scaling ${\cal Z}(t) \approx t^4$ and ${\cal J}(t) \approx t^{-8}$
for $t\ll t_*$. A nice choice for ${\cal J}(t)$ with this property (and satisfying ${\cal J}\simeq {\cal J}_0$ for $t\ll t_*$) is
\be
{\cal J}(t) = \frac{{\cal J}_0}{(1 + \frac{t}{t_*})^8}\,.
\ee
Equivalently, using~\eqref{piansatz},
\be
{\cal J}(\pi) =  \left(1 - e^{- (\pi-\pi_{\infty})}\right)^8 {\cal J}_0\,.
\label{jasfofpi}
\ee

Substituting ${\cal J}(t)$ into the integrated equation of motion~\eqref{rhocons} yields
\be
{\cal Z}(t) = - \left(f_{\infty}^2 + f_0^2\right)\frac{t^4}{t_*^4} + \left(1 + \frac{t}{t_*}\right)^4 f_0^2 \,,
\ee
where $f_{\infty}$, introduced for reasons that will soon become obvious, is related to the energy density by
\be
\rho =\frac{3}{2t_*^4}  \left(1 + \frac{f_{\infty}^2}{f_0^2}\right)\frac{f_0^3}{{\cal J}_0\Lambda^3}  \,.
\label{rhoexpres}
\ee
Moreover, we can obtain an expression for $\pi_{\infty}$ and the transition time:
\be
e^{\pi_{\infty}} = \sqrt{\frac{3f_0}{2{\cal J}_0\Lambda^3}} \frac{1}{|t_*|}\,.
\label{piinf}
\ee
In terms of $\pi$, the function ${\cal Z}$ can be expressed as
\be
{\cal Z}(\pi) = \frac{f_0^2}{\left(e^{\pi-\pi_{\infty}}-1\right)^{4}} \Bigg( e^{4 (\pi-\pi_{\infty})} - \left(1 + \frac{f_{\infty}^2}{f_0^2}\right)  \Bigg) \,.
\label{zasfofpi}
\ee
Hence we have 5 parameters defining the theory: $f_0$, $f_{\infty}$, $\Lambda$, ${\cal J}_0$ and $\pi_\infty$. The transition time $t_*$ is not a free parameter, as it is set by the other parameters in the Lagrangian. By construction, the Lagrangian~\eqref{piact} with the functions~\eqref{jasfofpi} and~\eqref{zasfofpi} admits the interpolating solution~\eqref{piansatz} as a solution to its equations of motion.

\subsection{Early times revisited}

With the expressions above for ${\cal Z}(\pi)$ and ${\cal J}(\pi)$, we can investigate the action at large, constant field values, where it takes the approximate form (note that this limit is essentially the early time limit on the solution~\eqref{piansatz}):
\be
{\cal L}_{\rm early} \simeq -\frac{f_{\infty}^2e^{2\pi_\infty}}{\left(e^{\pi-\pi_{\infty}}-1\right)^{4}}(\partial \pi)^2 + \frac{1}{{\cal J}_0}\frac{f_0^3}{2 \Lambda^3}\frac{(\partial \pi)^4}{ \left(1 - e^{- (\pi-\pi_{\infty})}\right)^8} \,,
\ee
It is convenient to define the almost-canonically-normalized variable,
\be
\phi = \frac{1}{1-e^{-(\pi-\pi_\infty)}}\,.
\label{phidef}
\ee
The virtue of this redefinition is that in terms of $\phi$, the background solution~\eqref{piansatz} reduces to a linear form
\be
\bar{\phi}(t) =  1 + \frac{t}{t_*}\,.
\label{phiback}
\ee
Another benefit is that the functions ${\cal Z}(\pi)$ and ${\cal J}(\pi)$ simplify to
\bea
\nonumber 
{\cal Z}(\phi) &=& f_0^2 \phi^4 - \left(f_0^2 + f_\infty^2 \right) \left(\phi - 1 \right)^4 \,;\\
{\cal J}(\phi) &=& \frac{{\cal J}_0}{\phi^8}\,.
\eea
Furthermore, the early-time action in terms of $\phi$ reduces to
\be
{\cal L}_{\rm early} \simeq - f_\infty^2e^{2\pi_\infty} (\partial\phi)^2 +\frac{f_0^3}{2{\cal J}_0\Lambda^3}(\partial\phi)^4 \,.
\label{Learlymain}
\ee
The kinetic term is healthy, as it should be, hence the theory admits Poincar\'e-invariant solutions. Further, the quartic term is manifestly {\it positive}: this ensures both a lack of superluminality about these Poincar\'e-invariant vacua and that the simplest dispersion relations following from S-matrix analyticity~\cite{Adams:2006sv} are satisfied.

\subsection{Radiative stability}

The reduced symmetry of the action due to the presence of the arbitrary functions ${\cal Z}(\pi)$ and ${\cal J}(\pi)$ makes it unlikely that the theory will be stable under quantum corrections. However, all is not lost. Recall that for large constant field values the action~\eqref{piact} can be cast as
\be
{\cal L}_{\rm early} \simeq f_\infty^2\left(-e^{2\pi_\infty} (\partial\phi)^2 +\frac{f_0^3}{2f_\infty^2{\cal J}_0\Lambda^3}(\partial\phi)^4\right)~.
\label{largefieldlag}
\ee
In this way, $f_\infty^2$ plays a role analogous to $1/\hbar$; for sufficiently large $f_\infty$, quantum effects can be made negligible and the theory will be radiatively stable~\cite{Nicolis:2009qm}.\footnote{Another way of saying this is that terms radiatively generated in the Lagrangian~\eqref{largefieldlag} will be suppressed by powers of $f_\infty$, and can be ignored for sufficiently large values of $f_\infty$.}

Similarly, at late times (or, as $\pi\to \infty$), the theory can be cast as
\be
{\cal L}_{\rm late} \simeq f_0^2 e^{2 \pi} (\partial \pi)^2 + \frac{f_0^3}{\Lambda^3}(\partial \pi)^2 \Box \pi + \frac{1}{{\cal J}_0}\frac{f_0^3}{2 \Lambda^3}(\partial \pi)^4~,
\ee
which is precisely of the form considered in~\cite{Creminelli:2012my}, where it was shown to be radiatively stable. Therefore, we see that the functional forms of ${\cal Z}(\pi)$ and ${\cal J}(\pi)$ are stable at both ends of the evolution. In between, most likely they will be greatly affected by quantum corrections. However, the fact that their asymptotic forms are preserved makes it plausible that a solution which interpolates between NEC-violating and NEC-satisfying regions will continue to exist. Although the detailed form of the solution will surely be modified, we do not expect its stability properties to be greatly affected, as we are able to satisfy the stability requirements for a wide range of parameters. In this sense the explicit interpolating form for ${\cal Z}(\pi)$ and ${\cal J}(\pi)$ constructed above is a proof of principle.

\section{NEC violation and neglecting gravity}
\label{gravneglect}

It is straightforward to calculate the stress-energy tensor for the Lagrangian~\eqref{piact}, in the approximation that the gravitational background is Minkowski space.
The energy density is constant and has already been given in~\eqref{rhocons} and~\eqref{rhoexpres}. The pressure is given by
\be
P = - {\cal Z}(\bar{\pi}) e^{2 \bar{\pi}} \dot{\bar{\pi}}^2 + \frac{1}{2{\cal J}(\bar{\pi})} \frac{f_0^3}{\Lambda^3} \dot{\bar{\pi}}^4 - 2 \frac{f_0^3}{\Lambda^3} \dot{\bar{\pi}}^2 \ddot{\bar{\pi}} \,.
\label{pressure}
\ee
On the solution~\eqref{piansatz}, at late times this reduces to\footnote{As a check, translating to the $\alpha$ parameter of Subluminal Genesis via~\eqref{jalpha}, this matches the pressure computed in~\cite{Creminelli:2012my}.}
\be
P_{\rm late} \simeq -\left(\frac{1}{{\cal J}_0}+2\right ) \frac{f_0^3}{\Lambda^3}\frac{1}{t^4}\,.
\label{Plate}
\ee
Since the late-time pressure grows as $1/t^4$ while the energy density remains constant, the NEC will violated at late times if $P_{\rm late} < 0$. This requires
\be
\frac{1}{{\cal J}_0} > -2 ~~~~~~~~~~~~({\rm NEC}~{\rm violation})\,.
\label{jNEC}
\ee
This is the NEC-violating genesis phase.

At early times, meanwhile, the pressure is constant and positive:
\be
P_{\rm early} \simeq \frac{1}{2t_*^4} \left(1+ 3\frac{f_\infty^2}{f_0^2}\right) \frac{f_0^3}{{\cal J}_0\Lambda^3}\,.
\ee
Hence the early-time regime is NEC-satisfying. 

\begin{figure}
\centering
\includegraphics[scale=0.4]{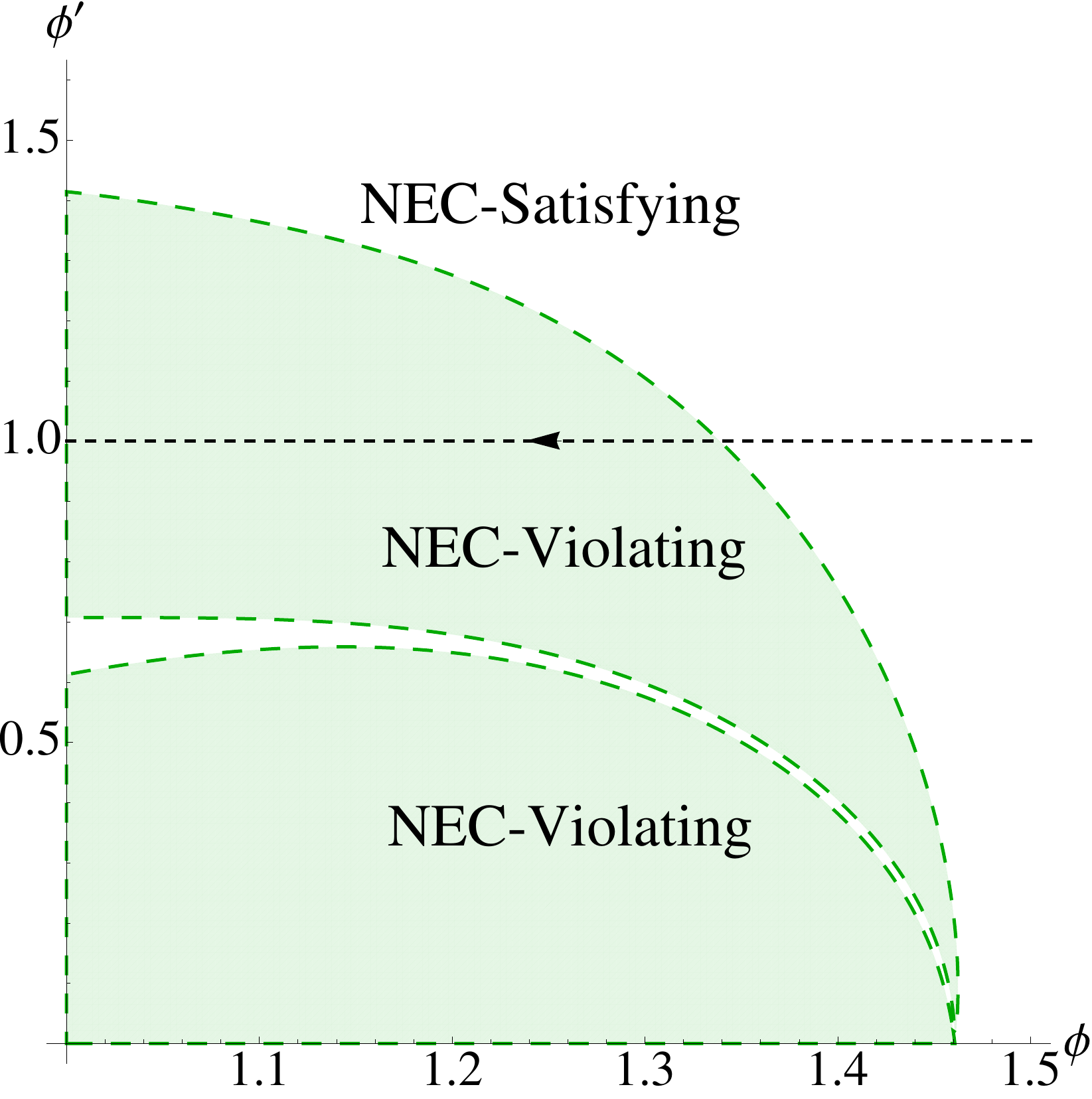}
\caption{\small NEC-satisfying and NEC-violating regions in the $(\phi,\phi')$ phase space, for the parameter values $\frac{f_{\infty}}{f_0} = 10$ and ${\cal J}_0 = 0.75$. The solution of interest, $\phi = 1+t/t_*$, corresponding to $\phi' = 1$, is plotted as a black dashed line. It first obeys the NEC for a period of time, and then crosses into the NEC-violating regime.}
\label{NECregion}
\end{figure}

More generally, by combining~\eqref{rhocons} and~\eqref{pressure} we see that the NEC is violated whenever
\be
\rho + P = 2\bigg(- {\cal Z}(\pi) e^{2\pi} \dot{\pi}^2 + \frac{1}{{\cal J}(\pi)} \frac{f_0^3}{\Lambda^3} \dot{\pi}^4 - \frac{f_0^3}{\Lambda^3} \dot{\pi}^2 \ddot{\pi}\bigg) < 0\,.
\label{NECvio}
\ee
where we have dropped the bars for simplicity. This condition can be studied numerically. For this purpose, we will focus on ``on-shell" solutions, that is, on profiles $\bar{\pi}(t)$ that are solutions to the equation of motion~(\ref{pieom1st}). This allows us to rewrite $\ddot{\pi}$ as a function of $\pi$ and $\dot{\pi}$. Moreover, it is convenient to express the result in terms of the $\phi$ variable introduced in~\eqref{phidef}, since its background evolution is particularly simple. The NEC-violating region in phase space corresponds to
\be
- \frac{3}{2} \frac{1}{f_0^2 {\cal J}_0} {\cal Z} (\phi) \phi^4 + \left(\frac{1}{{ \cal J}(\phi)} - 2\phi + 1 \right)\phi '^2 - 2  \frac{\phi'^2(\phi'^2 -1)}{1-2\phi'^2 -\left(1 + \frac{f_\infty^2}{f_0^2}\right) (1-\phi^{-1})^4} < 0~~~~~~~~({\rm NEC}~{\rm violation})\,,
\label{NECviol}
\ee
where we have defined $\phi' \equiv t_* \dot{\phi}$. The result is plotted in the $(\phi,\phi')$ plane in Fig.~\ref{NECregion} for a fiducial choice of parameters.

All of the results up to this point have been derived under the approximation that gravitational backreaction can be neglected. We will now quantify the time interval over which this
assumption is justified. Consider first the early-time regime. Since pressure and energy density are comparable (and constant) in this epoch, gravitational backreaction can
only be neglected for at most a Hubble time $H^{-1} = \sqrt{3M_{\rm Pl}^2/\rho}$. Our approximation is therefore justified for $t \gg t_{\rm beg}$, where (ignoring ${\cal O}(1)$ coefficients)
\be
\frac{t_{\rm beg}}{t_*} \sim \frac{1}{\sqrt{1 + \frac{f_\infty^2}{f_0^2}}} \sqrt{\frac{\Lambda^3}{f_0^3}} |t_*|M_{\rm Pl}\,.
\ee
For consistency, we must have $|t_{\rm beg}|\gg |t_*|$, that is,
\be
|t_*|\gg \frac{1}{M_{\rm Pl}} \sqrt{1 + \frac{f_\infty^2}{f_0^2}}\sqrt{\frac{f_0^3}{\Lambda^3}} \,.
\label{t*cond}
\ee
Determining the evolution before $t_{\rm beg}$, including gravity, would require a detailed calculation. But since the NEC is preserved, the answer is qualitatively simple: within a time of order $t_{\rm beg}$,
the evolution must trace back to a big bang singularity.

During the Genesis phase, on the other hand, the gravitational dynamics are dominated by the pressure~\eqref{Plate}. Integrating $M_{\rm Pl}^2\dot{H} \simeq -\frac{1}{2}P$,
we have
\be
H_{\rm late} \simeq \frac{1}{6M_{\rm Pl}^2|t|^3} (1+2{\cal J}_0) \frac{f_0^3}{{\cal J}_0\Lambda^3}\,,
\ee
corresponding to a time-dependent contribution to the energy density:
\be
\rho_{\rm late} = \frac{1}{12 M_{\rm Pl}^2t^6}  (1+2{\cal J}_0)^2\left( \frac{f_0^3}{{\cal J}_0\Lambda^3}\right)^2\,.
\ee
This dominates over the constant piece~\eqref{rhoexpres}. Gravitational backreaction can be neglected as long as $P_{\rm late}\gg \rho_{\rm late}$. 
This breaks down at a time $t_{\rm end}$ obtained by setting $P_{\rm late}\sim \rho_{\rm late}$:
\be
|t_{\rm end}| \sim \frac{1}{M_{\rm Pl}} \sqrt{\frac{f_0^3}{\Lambda^3}}\,.
\ee
The condition $f_0\gg \Lambda$ mentioned in~\eqref{flambcond} ensures that $|t_{\rm end}| \gg M_{\rm Pl}^{-1}$. 
Moreover, the condition~\eqref{t*cond} automatically implies that $|t_{\rm end}|\ll |t_*|$, which is obviously
required for consistency.

\begin{figure}
\centering
\includegraphics[width=5.5in]{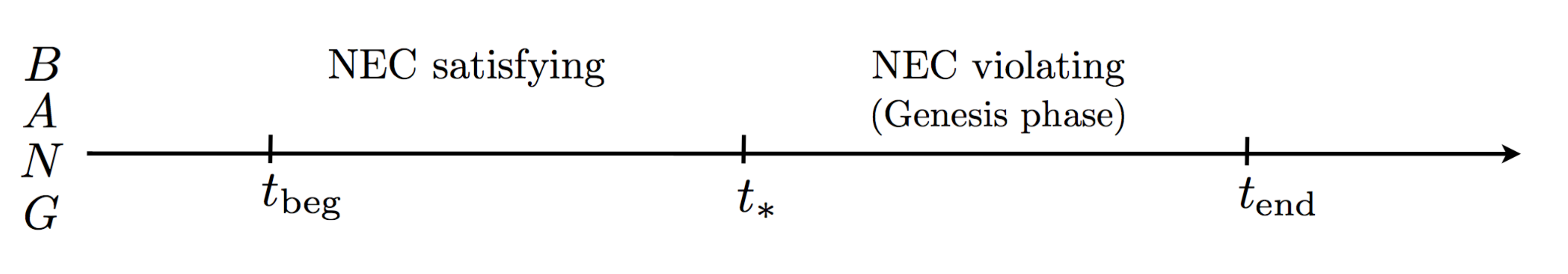}
\caption{\small Timeline for the evolution. Our approximation of neglecting gravity is valid for the range $t_{\rm beg} \leq t \leq t_{\rm end}$. For $t < t_{\rm beg}$, the universe asymptotes to a big bang singularity (since the NEC is satisfied in this regime).  At approximately $t_*$, the universe transitions from a NEC-satisfying phase to a NEC-violating one. For $t> t_{\rm end}$, cosmological expansion is important, and the universe must transition from the NEC-violating phase to a standard, radiation-dominated phase.}
\label{timeli}
\end{figure}

To summarize, our approximation of neglecting gravity is valid over the interval
\be
t_{\rm beg} \ll t \ll t_{\rm end}\,.
\ee
In order for the transition time to lie within this interval, $t_*$ must satisfy the condition~\eqref{t*cond}.
The time-line for the entire evolution is sketched in Fig.~\ref{timeli}. 

\subsection{Cosmological Evolution}
\label{cosmev}
\begin{figure}
\centering
\includegraphics[width=4.3in]{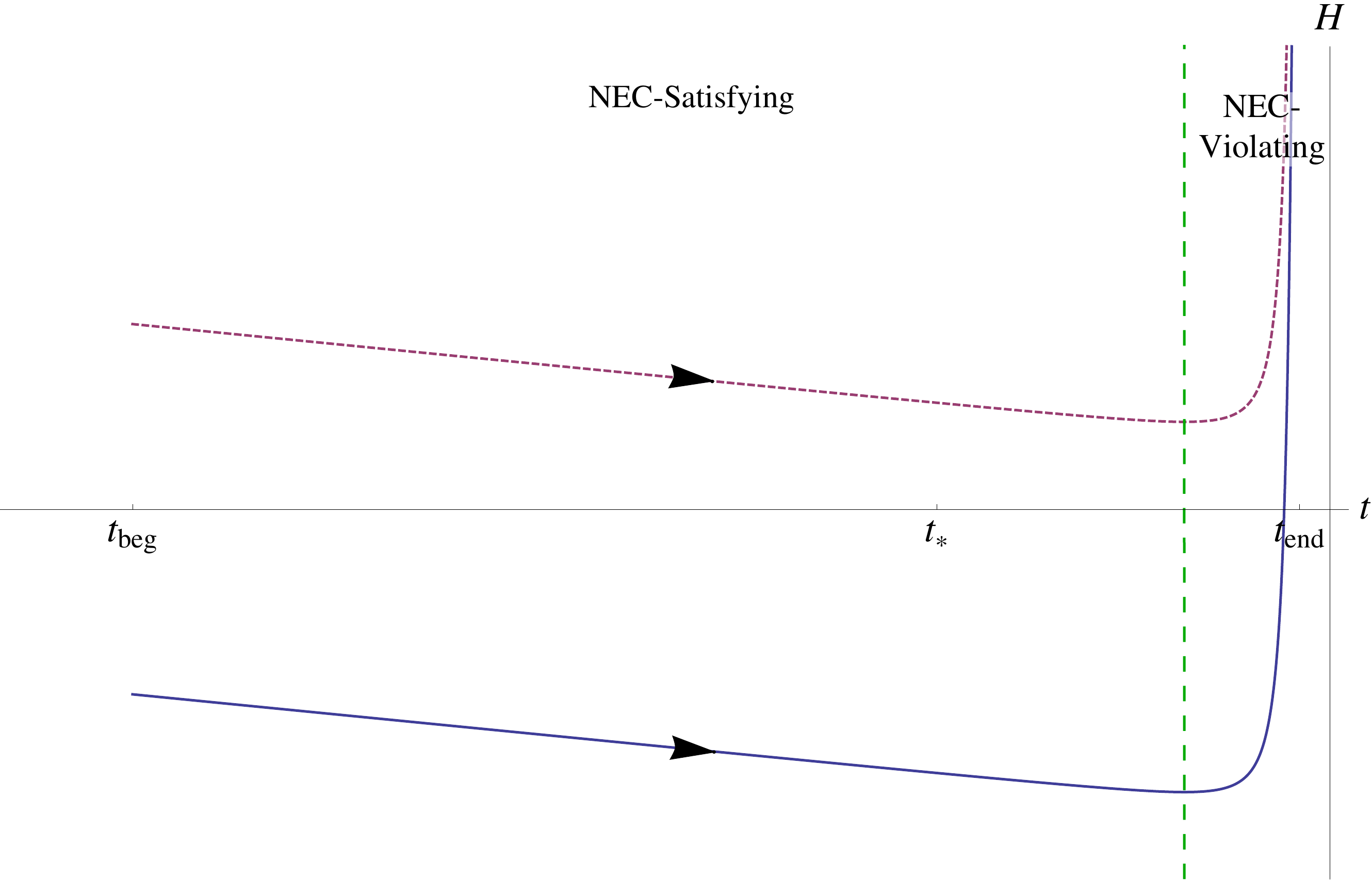}
\caption{\small The Hubble parameter is plotted for two values of $H_i$ with the fiducial parameters ${\cal J}_0 = 0.75$, $f_\infty/f_0 = 10$, $\Lambda  = 0.1$, $t_* = -1$, and $C = -6.3\times10^{-4} M_{\rm Pl}$.  The vertical dashed line marks the boundary between the NEC-satisfying and NEC-violating phases.  As we would expect, $\dot H < 0$ when the NEC holds, and $\dot H > 0$ when it is violated.  The dashed solution (corresponding to $H_i = 0.001M_{\rm Pl}$) always has $H>0$.  It represents an initially expanding universe with decelerating expansion, and could match onto a big-bang type solution for $t < t_{\rm beg}$. The solid line (with $H_i = - 0.001 M_{\rm Pl}$) represents an initially contracting universe $(H <0)$ which enters the phase of NEC-violation and undergoes a cosmological bounce to an accelerating phase $(H > 0)$, all within the regime of validity of our effective theory.}
\label{hubble_plot}
\end{figure}

With expressions for $P$, $\rho$, $t_{\rm beg}$, and $t_{\rm end}$, it is possible to show how the scale factor may smoothly transition from a decreasing phase to and increasing one.  Recall that the Hubble parameter obeys \eqref{Hdot}:
\be
M_{\rm Pl}^2\dot H = -\frac{1}{2}(\rho+P)~.
\ee
Combined with the expressions for $(\rho+P)$~\eqref{NECvio} and $e^{\pi(t)}$~\eqref{piansatz}, we obtain the following expression for $\dot H(t)$:
\be
\dot H(t) = \frac{-f_0^3}{M_{\rm Pl}^2 J_0 \Lambda^3 t_*^4} \left( \frac{3}{2} \left(1 + \frac{f_\infty^2}{f_0^2} \right)
- \frac{1}{2} \left(1 + \frac{t_*}{t} \right)^4 - J_0 \frac{t_*^4}{t^4} \frac{2 \frac{t}{t_*} + 1}{(1 + \frac{t}{t_*})^4} \right).
\label{Hdott}
\ee
This expression is valid in the range $t_{\rm end} \ll t \ll t_{\rm beg}$.  It may be integrated to give $H$ in this range:
\be
H(t) = \frac{-f_0^3}{M_{\rm Pl}^2 J_0 \Lambda^3 t_*^3} \left(\frac{3}{2} \left(1 + \frac{f_\infty^2}{f_0^2} \right)\frac{t}{t_*}
+ \frac{t_*^3}{6t^3} + \frac{t_*^2}{t^2} + \frac{3t_*}{t} - 2 \log{\frac{t}{t_*}} - \frac{t}{2t_*}
+ \frac{J_0}{3} \frac{t_*^3}{t^3} \frac{1}{(1 + \frac{t}{t_*})^3} \right) + C + H_i
\label{Hoft}
\ee
$C$ and $H_i$ are both integration constants, but $C$ is chosen to ensure that $H(t_{\rm beg}) = H_i$.
This is plotted over the range of validity in Fig.~\ref{hubble_plot} for two different values of $H_i$.  The dashed line solution represents an expanding universe that originated in a big bang, and the solid line is a contracting universe that might have originated from Minkowski space or a big bang singularity in the asymptotic past.  Although both solutions demonstrate the transition from $\dot H < 0$ to $\dot H > 0$, the solid line also shows that a solution can smoothly go from $H < 0$ to $H>0$ --- {\it i.e.}, a bounce. Notice that this bounce occurs before $t_{\rm end}$, our estimate for when gravitational back-reaction can no longer be ignored in the solution for $\pi$. This indicates that we can trust the existence of the bounce within our effective theory.

\section{Stability of perturbations}
\label{NECperts}

We now turn to the study of perturbations around background solutions. Expanding the Lagrangian~\eqref{piact} to quadratic order in perturbations $\varphi = \pi - \bar{\pi}$, we find
\be
{\cal L}_{\rm quad} = {\cal Z}_\varphi(t)\dot{\vp}^2 - {\cal K}_\varphi(t)  (\nabla \vp)^2 \,,
\label{vplag}
\ee
where we have defined the functions
\bea
\nonumber
{\cal Z}_\varphi(t) &\equiv& \(-{\cal Z}( \bar{\pi} ) e^{2 \bar{\pi} } + \frac{3}{{\cal J}( \bar{\pi} )} \frac{f_0^3}{\Lambda^3} \dot{ \bar{\pi} }^2\)\,; \\
{\cal K}_\varphi(t) &\equiv &  \(-{\cal Z}( \bar{\pi} ) e^{2 \bar{\pi} } + 2 \frac{f_0^3}{\Lambda^3} \ddot{ \bar{\pi} } + \frac{1}{{\cal J}( \bar{\pi} )} \frac{f_0^3}{\Lambda^3} \dot{ \bar{\pi} }^2\) \,.
\eea
The constraints on the quadratic theory are the following:

\begin{figure}
\centering
\subfigure[Gradient Instability]{%
\includegraphics[scale=0.33]{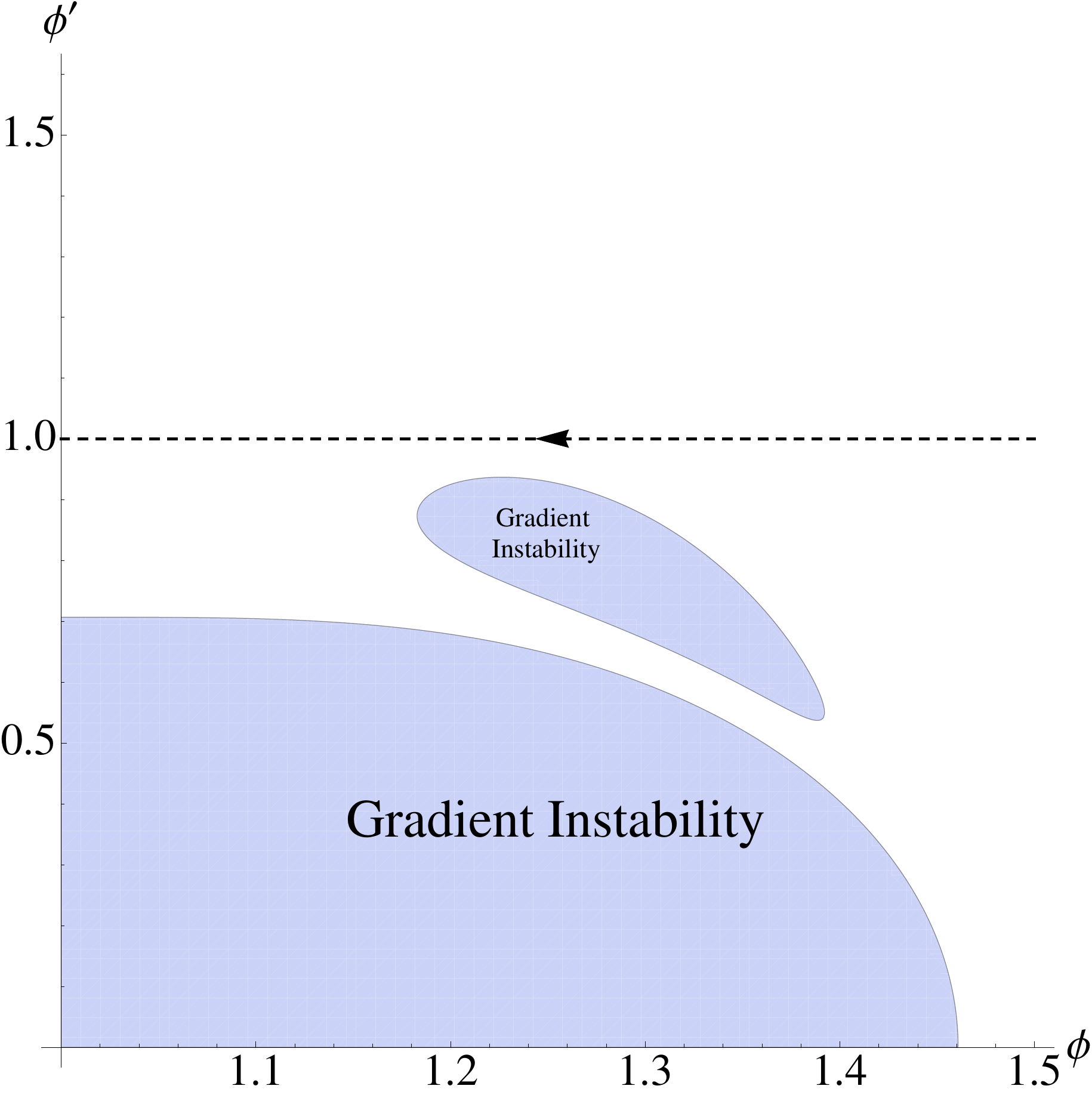}
\label{GRADregion}}
\subfigure[Superluminality]{%
\includegraphics[scale=0.35]{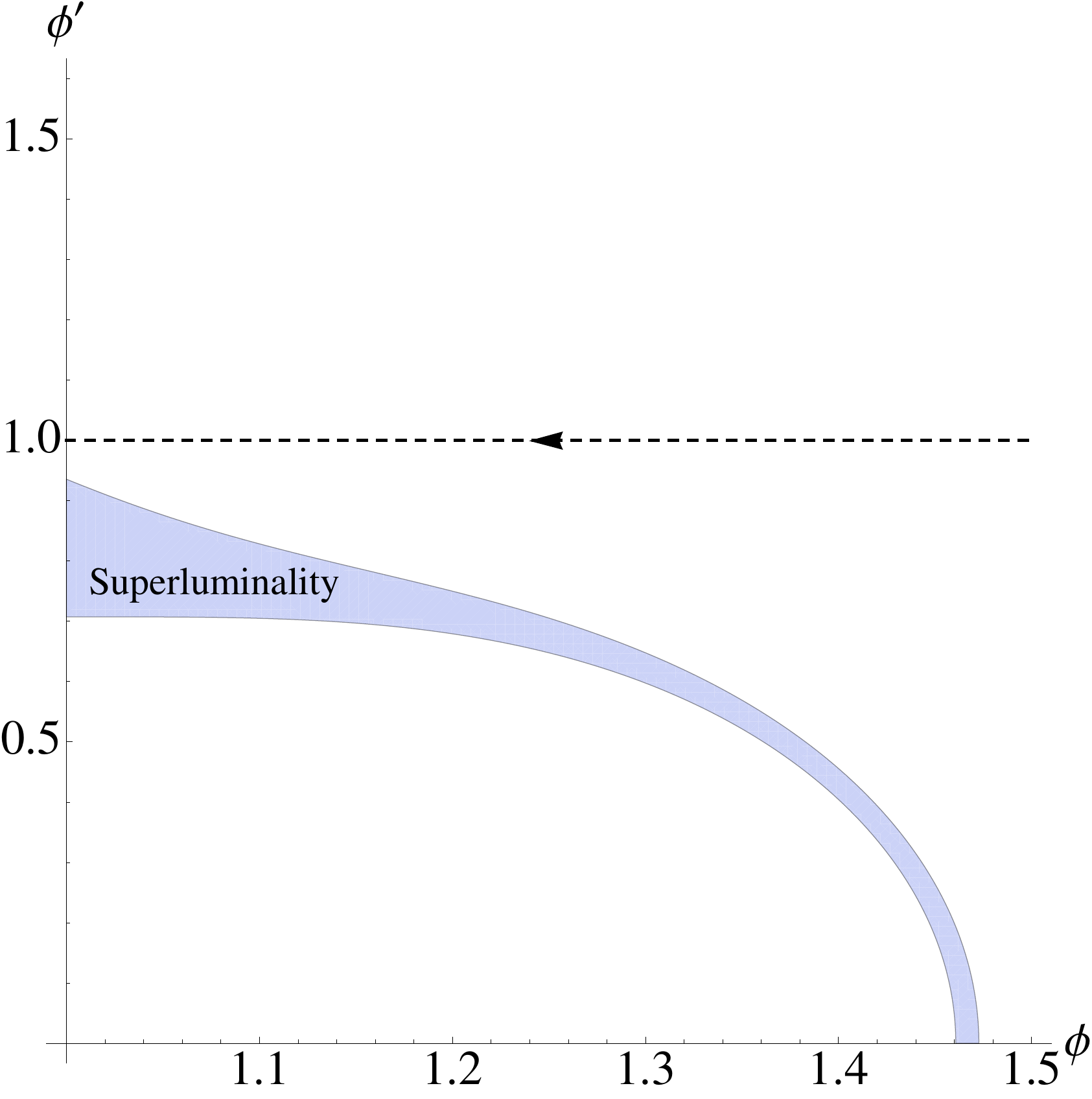}
\label{SUBLregion}}
\quad
\subfigure[Strong Coupling]{%
\includegraphics[scale=0.33]{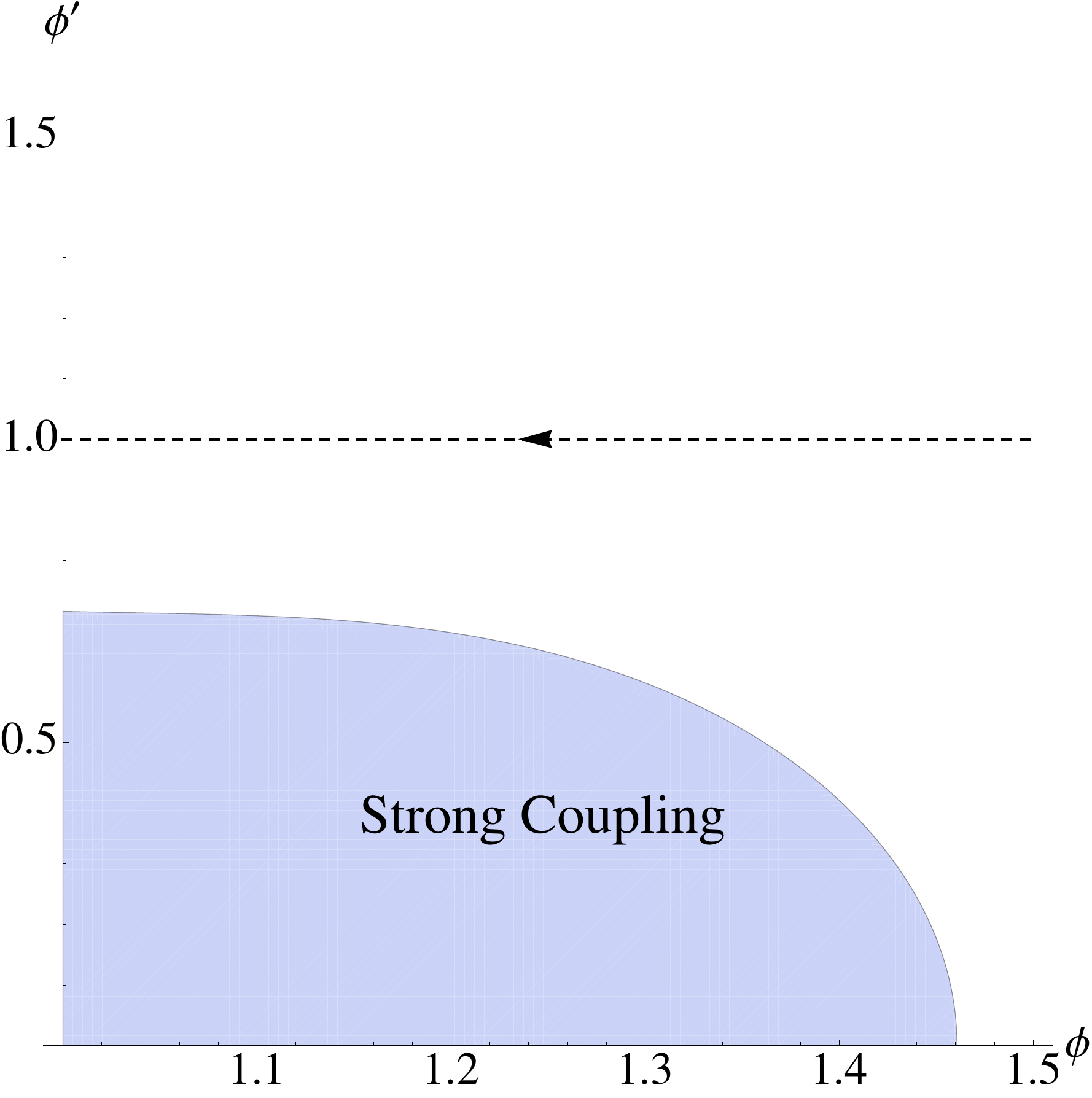}
\label{STRCregion}}
\caption{\small The shaded regions represent parts of the $(\phi,\phi')$ phase space where perturbations (a) suffer from gradient instabilities; (b) propagate superluminally; (c) are strongly coupled. The parameter values are $\frac{f_{\infty}}{f_0} = 10$ and ${\cal J}_0 = 0.75$. The solution of interest, $\phi = 1+t/t_*$, corresponding to $\phi' = 1$, is plotted as a black dashed line. It avoids all pathological regions.}
\label{constraints}
\end{figure}
\begin{itemize}

\item {\bf Absence of ghosts}: To avoid ghosts, the kinetic term should be positive: ${\cal Z}_\varphi > 0$. It is straightforward to show that will be the case if
\be
{\cal J}_0 > 0 ~~~~~~~~~~~~({\rm No}~{\rm ghosts})\,.
\label{noghost}
\ee
In particular, the NEC-violating condition~\eqref{jNEC} follows automatically.

\item {\bf Absence of gradient instabilities}: Similarly, the spatial gradient term should be positive: ${\cal K}_\varphi > 0$. Expressing this condition in terms of $\phi$, and dropping the bars for simplicity, we obtain
\bea
\nonumber
- \frac{3}{2}\frac{1}{f_0^2 {\cal J}_0}Z(\phi)\phi^4 + \left(2(2 \phi - 1) + \frac{1}{{\cal J} (\phi)} \right) \phi'^2  + 4 \frac{\phi'^2(\phi'^2 -1)}{1-2\phi'^2 -\left(1 + \frac{f_\infty^2}{f_0^2}\right) (1-\phi^{-1})^4} > 0 \\~~~~~~~~~~~({\rm No}~{\rm gradient}~{\rm instabilities})\,,
\label{nogradinst}
\eea
where we recall that $\phi' \equiv \dot{\phi}t_*$.

\item {\bf Subluminality}:  The final constraint at the quadratic level is to demand subluminal propagation: ${\cal K}_\varphi /{\cal Z}_\varphi < 1$. Assuming that both~\eqref{noghost} and~\eqref{nogradinst} are satisfied, subluminality follows by definition if the kinetic term is larger than the gradient term. It is straightforward to show that this will be the case if
\be
\left( \frac{1}{{\cal J}(\phi)} - 2 \phi + 1 \right) \phi'^2  - 2  \frac{\phi'^2(\phi'^2 -1)}{1-2\phi'^2 -\left(1 + \frac{f_\infty^2}{f_0^2}\right) (1-\phi^{-1})^4} > 0 ~~~~~~~~~~~({\rm Subluminal})\,.
\label{sublumcond}
\ee
In the genesis regime (corresponding to $\phi \rightarrow 1$ and $\phi' \rightarrow 1$), in particular, this gives a constraint on the constant ${\cal J}_0$:
\be
{\cal J}_0 < 1\,.
\label{J0cond}
\ee
\end{itemize}

Beyond the quadratic theory, we should also check that the interactions are perturbative. Consider the cubic vertex for the perturbations:
\be
{\cal L}_3 = \frac{f_0^3}{\Lambda^3} (\partial\varphi)^2\Box\varphi\,,
\ee
After canonical normalization of the kinetic term, $\varphi_c \equiv {\cal Z}_\varphi^{1/2} \varphi$, the cubic term becomes suppressed by the effective strong coupling scale
\be
\Lambda_{\rm eff}  = \frac{\Lambda}{f_0} {\cal Z}_\varphi^{1/2}\,.
\ee
For consistency of the effective field theory, the characteristic frequency of the background, namely $\dot{\bar{\pi}}$, should lie below this cutoff: 
\be
\dot{\bar{\pi}} \ll \Lambda_{\rm eff}\,. 
\label{lambeff}
\ee
In the Genesis phase, in particular, $\dot{\bar{\pi}} \simeq 1/t$ sets the scale at which perturbations freeze out. Hence~\eqref{lambeff} is necessary to consistently describe the generation of perturbations within the effective theory. A straightforward calculation shows that this condition implies:
\be
 2 \bar{\phi}'^2 \left(\frac{{\cal J}_0 \Lambda}{3 f_0 \bar{\phi}^8} - 1\right)\ll \left (1 + \frac{f_{\infty}^2}{f_0^2}\right) \left(1 - \bar{\phi}^{-1}\right)^4 - 1 ~~~~~~~({\rm weak}~{\rm coupling})\,.
\label{noSC}
\ee
Note that the left-hand side is negative-definite within the range $0 < {\cal J}_0 < 1$ allowed by~\eqref{noghost} and~\eqref{J0cond}.  As a check,  note that in the genesis regime,~\eqref{noSC} reduces to $\frac{{\cal J}_0\Lambda}{f_0} \ll \frac{3}{2}$, which is another way to confirm the condition $f_0\gg \Lambda$ mentioned in~\eqref{flambcond}.

In summary, the allowed range of ${\cal J}_0$ values is
\be
0 < {\cal J}_0 < 1\,.
\ee
The remaining constraints --- no gradient instabilities~\eqref{nogradinst}, subluminality~\eqref{J0cond}, and weak coupling~\eqref{noSC} --- are plotted in the $(\phi,\phi')$ phase space in Fig.~\ref{constraints} for a fiducial choice of parameters.  

\begin{figure}
\centering
\includegraphics[scale=0.5]{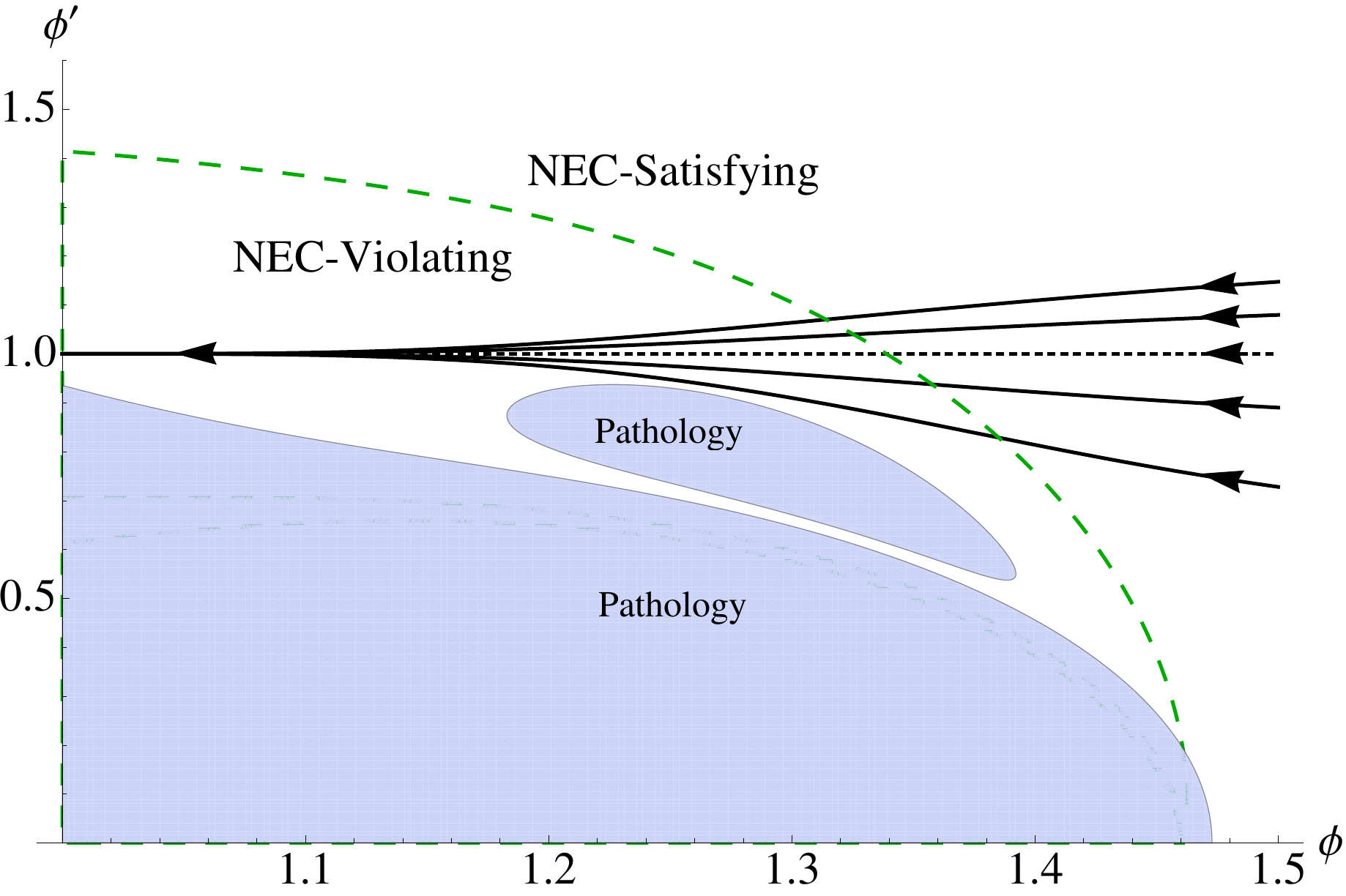}
\caption{\small Phase portrait with all constraints overlaid, again for the fiducial choice of parameters $\frac{f_{\infty}}{f_0} = 10$ and ${\cal J}_0 = 0.75$. The shaded region represents the union of all pathological regions shown in Fig.~\ref{constraints}. The green long-dashed line separates the NEC-satisfying and NEC-violating regions. The black short-dashed line corresponds to the background solution of interest, given by~\eqref{phiback}. The solid lines represent other background solutions (with different initial conditions).}
\label{sumcons}
\end{figure}

These constraints are overlaid in Fig.~\ref{sumcons} with a range of solutions to the equation of motion.  On these plots, the background solution~\eqref{phiback} of interest corresponds to $\phi' = 1$. Other background solutions, corresponding to different initial conditions, are also plotted as solid lines. This shows that there is a wide range of trajectories that interpolate between a constant field profile at early times and the Genesis solutions at late times, while avoiding the pathological region at all times. Furthermore, it is clear that the background solution $\phi' = 1$ is an attractor at late times.

\section{Conclusions}
\label{conclu}

It has proven surprisingly difficult to violate the null energy condition with a well-behaved relativistic quantum field theory. In the simplest attempts, violating the NEC generally introduces ghost instabilities, gradient instabilities, superluminality, or absence of a Lorentz-invariant vacuum. Progress has been made in avoiding some of these shortcomings, but a fully satisfactory example remains to be found. The null energy condition appears to be connected to some fundamental physics principles, such as black hole thermodynamics and the (non)-existence of cosmological bounces. Therefore, if it turns out that violating the NEC is impossible, pinpointing which of the aforementioned pathologies is the real roadblock will tell us something fundamental.

The recently-proposed DBI Genesis scenario is the first example of a theory admitting both a Poincar\'e-invariant vacuum and NEC-violating solutions. As argued by Rubakov, however, these two backgrounds lie on different branches of solutions and cannot be connected by a smooth solution without strong coupling occurring. This is an immediate consequence of dilation invariance.

Here, we have abandoned dilation symmetry in order to circumvent Rubakov's no-go argument. We have constructed a theory which admits a time-dependent solution that smoothly interpolates between a NEC-satisfying phase at early times and a NEC-violating phase at late times. There exists a wide range of parameters for which perturbations around the background are stable, comfortably subluminal and weakly-coupled at all times.

The main drawback of the construction is the presence of suitably-engineered interpolating functions in the action. It is highly unlikely that the detailed form of these functions will be preserved by quantum corrections. However, we argued that their asymptotic forms both in the past and in the future are radiatively stable. Moreover, our analysis did not depend sensitively on the details of the interpolation. Therefore, all we need is for the quantum-corrected action to still allow an interpolation between NEC-satisfying and NEC-violating solutions. We leave a detailed analysis of radiative stability to future work.

Another drawback of the explicit example presented here is that the kinetic term flips sign as we adiabatically vary $\phi$. It is healthy at early times, consistent with Poincar\'e invariance, but becomes
ghost-like at late times, which is necessary to obtain a NEC-violating solution with the cubic Galilean Genesis action. Of course, as mentioned earlier, perturbations around the time-dependent
background are always healthy. However, it would be aesthetically desirable if the perturbations around $\phi = {\rm constant}$ backgrounds were also healthy for all field values of interest. This should be achievable
by deforming the DBI Genesis Lagrangian, since this theory precisely satisfies this property while allowing a NEC-violating background. We plan to study the DBI Genesis generalization in the future. It is also possible that the DBI extension will alleviate the quantum stability issues discussed above.

{\bf Acknowledgements:} We would like to thank Lasha Berezhiani and Kurt Hinterbichler for useful discussions. This work is supported in part by NASA ATP grant NNX11AI95G, NSF CAREER Award PHY-1145525 (B.E. and J.K.); the Kavli Institute for Cosmological Physics at the University of Chicago through grant NSF PHY-1125897 and an endowment from the Kavli Foundation and its founder Fred Kavli and by the Robert R. McCormick Postdoctoral Fellowship (A.J.).

\bibliographystyle{utphys}
\addcontentsline{toc}{section}{References}
\bibliography{necbib}

\end{document}